\documentclass[12pt]{iopart}

\expandafter\let\csname equation*\endcsname\relax
\expandafter\let\csname endequation*\endcsname\relax

\usepackage[a4paper,margin=3cm,bottom=3.5cm]{geometry}

\usepackage{amssymb,amsthm,amsmath,mathtools} 
\usepackage{iopams}  
\usepackage{color}
\usepackage{xcolor}

\usepackage{braket}
\usepackage{enumitem}
\usepackage[unicode]{hyperref}

\usepackage{caption}            
\usepackage{subcaption}
\usepackage{wrapfig}

\usepackage[
  sorting=none,
  backend=bibtex,
  url=false,
  style=numeric]{biblatex}
  
\theoremstyle{remark}
\newtheoremstyle{remarkstyle}
  {\topsep}
  {\topsep}
  {\addtolength{\leftskip}{1em}\addtolength{\rightskip}{1em} \itshape}
  {-0.8em}
  {\slshape}
  {---}
  { }
  {\thmname{#1} \thmnote{(#3)}} 

\theoremstyle{remarkstyle}
\newtheorem*{remark}{Remark}

\renewcommand{\Re}{\mathrm{Re}}

\newcommand{\R}{\mathbb{R}} 
\newcommand{\C}{\mathbb{C}} 

\newcommand{\gateaux}{\mathfrak{D}}
\newcommand{\chamil}{\mathfrak{H}}
\newcommand{\Iden}{\mathrm{I}}

\newcommand{\di}{\mathrm{i}}
\newcommand{\dd}{\mathrm{d}}
\newcommand{\de}{\mathrm{e}}

\newcommand{\norm}[1]{\left\| #1 \right\|}
\newcommand{\abs}[1]{\left| #1 \right|}

\newcommand{\supnorm}{\mathrm{sup}} 

\begin{document}

\title[Robustness of optimal quantum annealing protocols]{Robustness of optimal quantum annealing protocols}

\author{Niklas Funcke$^{1,2}$, Julian Berberich$^1$}

\address{$^1$University of Stuttgart, Institute for Systems Theory and Automatic Control, 70569 Stuttgart, Germany}
\address{$^2$corresponding author}

\ead{niklas.funcke@ist.uni-stuttgart.de, \hspace*{1.1cm} julian.berberich@ist.uni-stuttgart.de}

\vspace{10pt}
\begin{indented}
\item[]\today
\end{indented}

\begin{abstract}
 Noise in quantum computing devices poses a key challenge in their realization. 
 In this paper, we study the robustness of optimal quantum annealing protocols against coherent control errors, which are multiplicative Hamiltonian errors causing detrimental effects on current quantum devices. We show that the norm of the Hamiltonian quantifies the robustness against these errors, motivating the introduction of an additional regularization term in the cost function. We analyze the optimality conditions of the resulting robust quantum optimal control problem based on Pontryagin’s maximum principle, showing that robust protocols admit larger smooth annealing sections.
 This suggests that quantum annealing admits improved robustness in comparison to bang-bang solutions such as the quantum approximate optimization algorithm.
 Finally, we perform numerical simulations to verify our analytical results and demonstrate the improved robustness of the proposed approach.
\end{abstract}

\vspace{2pc}
%
%


\section{Introduction}

Quantum computing has the potential to solve certain computational problems faster than classically possible~\cite{grover1996fast,shor1997polynomial,nielsen_quantum_2010}. A frequently considered problem on current noisy intermediate-scale quantum (NISQ,~\cite{preskill_quantum_2018,bharti2021nisq}) devices is to steer a quantum state to the ground state of a given cost Hamiltonian~\cite{sanders2020compilation,abbas2023quantum,gemeinhardt2023quantum}. Both quantum annealing (QA,~\cite{kadowaki_quantum_1998,farhi2000quantum}) and the quantum approximate optimization algorithm (QAOA,~\cite{farhi_quantum_2014}) tackle this problem by interpolating or iterating between two Hamiltonians.

The interface of variational quantum algorithms and quantum optimal control has received significant attention in recent years~\cite{magann_pulses_2021,koch2022quantum}. In particular, various recent works have used Pontryagin's maximum principle~\cite{pontryagin1962mathematical,kirk_optimal_2004,liberzon_calculus_2012} to study optimality of quantum algorithms, compare~\cite{yang_optimizing_2017,lin2019application,boscain_introduction_2021,brady_optimal_2021} and references therein. 
In the context of quantum optimization,~\cite{yang_optimizing_2017} showed that a bang-bang structure, which alternates between two Hamiltonians as in QAOA, produces optimal results.
However, it was later shown that, in general, the optimal solution contains singular sections with smoothly varying inputs as in QA~\cite{brady_optimal_2021}.

In the current NISQ era, noise poses a key challenge for experimental realizations of quantum devices~\cite{preskill_quantum_2018,bharti2021nisq}. Quantum errors can be categorized into incoherent~\cite{nielsen_quantum_2010} and coherent~\cite{kaufmann2023characterization} errors, where the latter were found to be particularly detrimental for quantum error correction schemes~\cite{sanders2016bounding,bravyi2018correcting,ouyang2021avoiding}.

In the present paper, we investigate the robustness of optimal QA protocols against one important class of coherent errors: coherent control errors. These errors can be caused by miscalibration due to imprecise classical control, leading to possible over- or underrotations of quantum gates. Various studies have shown that, among the class of coherent errors, coherent \emph{control} errors are especially crucial on current quantum hardware~\cite{barnes_quantum_2017,trout2018simulating,arute2019quantum}. With this motivation, different techniques have been developed to cope with coherent control errors, e.g., composite pulses~\cite{levitt_composite_1986}, dynamically error-corrected gates~\cite{khodjasteh_dynamically_2009}, quantum error correction~\cite{bravyi_correcting_2018,debroy_stabilizer_2018}, randomized compiling~\cite{wallman_noise_2016} or hidden inverses~\cite{zhang_hidden_2022}.

In this paper, we study the robustness of optimal QA protocols by deriving a Lipschitz bound, which shows that robustness against coherent control errors can be quantified via the norm of the Hamiltonian. This result generalizes previous findings on Lipschitz bounds for coherent control errors, which were derived in~\cite{berberich_robustness_2023} for a simpler setup with unitary gates affected by constant errors.
We then propose a \emph{robust} QA protocol which includes an additional regularization term penalizing the norm of the Hamiltonian.
This regularization encourages optimal solutions which are inherently more robust and, therefore, produce more reliable results in the presence of coherent control errors.
Next, we study the optimal solution of the resulting robust quantum optimal control problem based on Pontryagin's maximum principle, showing that it is substantially different with a larger smooth annealing section in comparison to the nominal (i.e., not robust) setup from~\cite{brady_optimal_2021}.
This indicates that smoothly varying Hamiltonians as in QA admit improved robustness properties compared to pure bang-bang solutions which arise in QAOA, compare~\cite{yang_optimizing_2017,brady_optimal_2021}.
Finally, we demonstrate that the proposed robust QA protocol indeed outperforms existing approaches without regularization in simulations with coherent control errors.
While existing approaches to cope with noise in QA rely on explicit correction steps~\cite{pudenz2015quantum,pearson2019analog}, our findings show the potential of inherent robustness properties which may simplify implementations on noisy devices by following a more robust design.

The remainder of the paper is structured as follows. First, in Section~\ref{sec:quantum_annealing}, we introduce the basic problem. In Section~\ref{sec:robust_annealing}, we derive a Lipschitz bound which quantifies robustness based on the norm of the Hamiltonian and we use it to state a robust QA protocol.
The optimal solution of this protocol is studied in Section~\ref{sec:robust_optimal_annealing} based on Pontryagin's maximum principle.
Section~\ref{sec:simulations} contains numerical results confirming the preceding theoretical analysis.
Finally, the paper is concluded in Section~\ref{sec:conclusion}.
Additionally, the appendix contains technical assumptions and proofs of parts of the main results.

\section{Quantum annealing and Pontryagin’s maximum principle}\label{sec:quantum_annealing}

QA can be used to solve combinatorial problems, e.g., by formulating them as Ising models~\cite{kadowaki_quantum_1998, lucas_ising_2014}. The main idea is to find a quantum state $\ket{x}$ which minimizes a quadratic form of a problem-dependent Hamiltonian matrix $C$, e.g.,  
\begin{align}
	J \coloneq \min_{\norm{\ket{x}}_2 = 1} \quad \braket{x |C| x}. \label{eqs:objectivefunction}
\end{align}

The ground state encodes the solution of the computational problem~\cite{lucas_ising_2014}. 

In QA, the protocol $u$ is found by steering the quantum system towards the ground state by smoothly varying a parametrized Hamiltonian $H(u)$.
In addition, due to physical limitations, the annealing protocols may need to satisfy bounds. Such problems can be formulated as optimal control problems and analyzed by using Pontryagin's maximum principle~\cite{yang_optimizing_2017, brady_optimal_2021}.

To introduce QA, we denote $C$ as the problem Hamiltonian and $B$ as the mixer Hamiltonian. In the context of Ising models, they are defined using the Pauli matrices $\sigma^x,\sigma^z$ as
\begin{align}
	B = - \sum_{i=1}^N \sigma_i^x, \qquad C = \sum_{i,j = 1}^N J_{ij} \sigma_i^z \sigma_j^z. \label{eqs:isingmodel}
\end{align}

\paragraph{Optimal quantum annealing.}\label{prb:optimalcontrol}
  Given $H(u)$, find the optimal annealing protocol $u : [0,T] \to [0,1]$ with respect to the cost $\braket{x(T) |C| x(T)}$, i.e.,
\begin{equation}
	\begin{aligned}
	  & \min_{u(\cdot),\ket{x(\cdot)}} \quad  \braket{x(T)|C|x(T)} \\ 
	  & \begin{aligned}
        \quad\mathrm{\textnormal{s.t.}} \quad& \ket{\dot{x}} = - \di H(u) \ket{x(\tau)} \qquad & \forall_{\tau \in [0,T]} \\
			 & \ket{x(0)} = \ket{x^B} \\
			 & \;\, u(\tau) \in [0,1] & \forall_{\tau \in [0,T]},
      \end{aligned}\label{eqs:prboptimalcontrol}
	\end{aligned}
\end{equation}
where $\ket{x^B}$ is the ground state of $B$ and
\begin{align}
     H(u) &= u B + (1-u) C.
\end{align}

The idea of QA is based on the adiabatic theorem~\cite{farhi_quantum_2000}. A practical and widespread class of possible solutions for the above problem consists of bang-bang control strategies, commonly referred to as the quantum approximate optimization algorithm (QAOA,~\cite{farhi_quantum_2014}). Here, either the Hamiltonian $B$ or the Hamiltonian $C$ is applied to the quantum system in an alternating fashion. Bang-bang protocols are motivated by their optimality in different optimal control setups~\cite{kirk_optimal_2004}.

Using Pontryagin's maximum principle, previous studies showed that optimal QA protocols can contain singular sections~\cite{brady_optimal_2021}. A singular section is a section in the QA protocol where $u$ lies in the interior of $[0,1]$, i.e., a Hamiltonian that smoothly interpolates between $B$ and $C$ is applied to the quantum system. In addition,~\cite{brady_optimal_2021} shows that optimal QA protocols, i.e., solutions of the above QA problem, always start with a bang and always end with a bang. This has led to new insights into the design of optimal QA protocols. 

\section{Robust optimal quantum annealing protocols}\label{sec:robust_annealing}

In practice, noise poses a key challenge in quantum computing. While previous results on optimal QA protocols~\cite{yang_optimizing_2017,brady_optimal_2021} focused on a noise-free quantum state evolution, in the presence of noise, the optimal solution might be substantially different.

In the following, we propose a \emph{robust} optimal annealing protocol which leads to a solution leading to a small cost even in the presence of noise.
We consider coherent control errors in this paper. Coherent control errors denote the class of perturbations which can be described as a multiplicative noisy term in the Hamiltonian.

\paragraph{Coherent control error.}\label{def:coherentcontrolerr}
 A coherent control error is a perturbation signal $\epsilon(\tau)$ such that the quantum system evolves according to the following noise Schrödinger equation
  \begin{align}
    \ket{\dot{x}_\epsilon(\tau)} = - \di (1+\epsilon(\tau)) H(u(\tau)) \ket{x_\epsilon(\tau)} \label{eqs:coherentcontrolerr}.
  \end{align}

Throughout this paper, we assume that the error signal is bounded as $\abs{\epsilon(\tau)} \leq \hat{\epsilon}$ for all $\tau \in [0,T]$, and we assume that the solution exists and is unique. See~\ref{apx:assumptions} for the technical assumptions required for the following analysis.

Our theoretical results involve the concept of a Lipschitz bound, which quantifies robustness by bounding the worst-case difference between solutions of the differential equation~\eqref{eqs:coherentcontrolerr} for different error terms.

\paragraph{Lipschitz bound on fidelity~\cite{berberich_robustness_2023}.}\label{lem:lipboundfidelity}
 Suppose $L$ is a Lipschitz bound of $\ket{x_\epsilon}$, i.e.,
\begin{align}
  \norm{ \ket{x_{\epsilon'}(T)} - \ket{x_\epsilon (T)} }_2 \leq L \sup_{\tau \in [0,T]} \abs{ \epsilon'(\tau) - \epsilon(\tau) } = L \norm{ \epsilon' - \epsilon }_{\supnorm}, \label{eqs:deflipschitz}
\end{align}
where $\epsilon,\epsilon' : [0,T] \to \R$ are error signals. Then, for any $\epsilon : [0,T] \to \R$ satisfying $\forall_{\tau \in [0,T]} \; \abs{\epsilon(\tau)} \leq \hat{\epsilon}$ for some $\hat{\epsilon} \geq 0$ and any initial condition $\ket{x^0}$, it holds that
\begin{align}
  \abs{\braket{x_\epsilon(T) | x_0(T)}} \geq 1 - \frac{L^2 \hat{\epsilon}^2}{2}. \label{eqs:lipboundfidelity}
\end{align}

Note here that $\sup_{\tau \in [0,T]} \abs{\epsilon'(\tau)-\epsilon(\tau)}$ is a finite number by assumption. Thus, if a Lipschitz bound as in~\eqref{eqs:deflipschitz} is available, the fidelity decrease can be bounded by $\frac{L^2 \hat{\epsilon}^2 }{2}$. A proof of this statement based on~\cite{berberich_robustness_2023} can be found in the~\ref{apx:thmrobustbound}. To summarize, the worst-case fidelity can be bounded by two components: the error bound $\hat{\epsilon}$ and the Lipschitz bound $L$. This motivates the derivation of a Lipschitz bound in the following.

\paragraph{Lipschitz bound.} A Lipschitz bound as in~\eqref{eqs:deflipschitz} can be computed as the integral over the norm of the Hamiltonian, i.e.,
\begin{align}
	L = \int_{0}^{T} \norm{H(u(\tau))}_2 \;\dd \tau \label{eqs:lipschitzupperbound},
\end{align}
where $\norm{\cdot}_2$ denotes the spectral norm. The proof is given in~\ref{apx:thmlipschitzbound} and requires technical assumptions on boundedness and uniqueness of solutions, compare~\ref{apx:assumptions}. In view of the fidelity bound $\frac{L^2 \hat{\epsilon}^2}{2}$, this means that a smaller norm of the Hamiltonian implies an improved robustness against coherent control errors. While an analogous result was derived in~\cite{berberich_robustness_2023} for the robustness of unitary quantum gates, our bound applies to time-varying errors $\epsilon(\tau)$ affecting the continuous-time evolution of the state $\ket{x(\tau)}$ in~\eqref{eqs:coherentcontrolerr}. This poses several additional technical challenges, compare~\ref{apx:thmlipschitzbound}.

\begin{remark}
  Due to the equivalence of norms, also different matrix norms than the spectral norm can be used in~\eqref{eqs:lipschitzupperbound}, e.g., the Frobenius norm which is differentiable everywhere except at zero.
\end{remark}

To reflect the need to reject control errors in the ground state preparation, we propose a slightly modified optimal control problem, which includes a robustness measure in the objective function. 

\paragraph{Robust optimal quantum annealing.}\label{prb:roboptimalcontrol}
  Given $H(u)$, find the optimal annealing protocol $u : [0,T] \to [0,1]$ with respect to the cost $\braket{x(T) |C| x(T)}$, i.e.,
  \begin{align}
      \min_{u(\cdot),\ket{x(\cdot)}} \quad&  \braket{x(T)|C|x(T)}  + \zeta \int_{0}^{T} \norm{H(u(\tau))}_2 \;\dd \tau \\
      \quad\mathrm{\textnormal{s.t.}} \quad& \textrm{(constraints in equation (\ref{eqs:prboptimalcontrol}))}.
    \label{eqs:prbrobustoptimalcontrol}
  \end{align}

The parameter $\zeta \geq 0$ denotes the weighting factor between optimal preparation of the ground state and robustness against coherent control errors. 
One can view the robustness measure as a regularization which trades off an optimal solution of the computational problem and robustness against coherent control errors. 

\section{Optimal solution of robust quantum annealing}\label{sec:robust_optimal_annealing}

In the following, we use Pontryagin's maximum principle~\cite{kirk_optimal_2004,liberzon_calculus_2012} to study optimal solutions of the robust optimal QA problem.

We show that the solution is different from the nominal quantum optimal control problem in Section~\ref{prb:optimalcontrol} with a significantly larger singular region and, thus, a larger annealing section.
A singular control describes those portions of the optimal protocol $u^\ast$ that lie in the interior of the input constraint set $[0,1]$.

In the following, we will omit the time argument of the involved variables, except at selected places. The theoretical analysis relies on the control Hamiltonian $\chamil$ with co-state $\ket{\lambda}$ defined by
\begin{align}
  \chamil(\ket{x},\ket{\lambda},u)
    &= \underbrace{\di \braket{x | H(u) | \lambda} - \di \braket{\lambda | H(u) | x}}_{h(\ket{x},\ket{\lambda},u)} - \zeta q(u) \\
    &= h(\ket{x},\ket{\lambda},u) - \zeta q(u).
\end{align}
where $q(u) = \norm{H(u)}_2$. According to Pontryagin's maximum principle~\cite{kirk_optimal_2004,liberzon_calculus_2012}, the optimal solution $u^\ast$ of the robust quantum optimal control problem in Section~\ref{prb:roboptimalcontrol} satisfies
\begin{align}
  \chamil(\ket{x^\ast},\ket{\lambda^\ast},u^\ast) \geq \chamil(\ket{x^\ast},\ket{\lambda^\ast},u) & \qquad \forall_{\tau \in [0,T]}, \; \forall_{u \in [0,1]}, \label{eqs:pmpmaximumprinciple}
\end{align}
where $\ket{x^\ast}$, $\ket{\lambda^\ast}$ and $u^\ast$ denote the optimal trajectories. Using equation~\eqref{eqs:pmpmaximumprinciple}, the optimal protocol $u^\ast(\tau)$ can be found through the following optimization problem
\begin{align}
    u^\ast(\tau) &\in \arg \max_{u \in [0,1]} h(\ket{x^\ast(\tau)},\ket{\lambda^\ast(\tau)},u) - \zeta q(u)  \qquad \forall_{\tau \in [0,T]}. \label{eqs:optprbustar}
\end{align}

It follows from the Weierstrass extreme value theorem~\cite{keisler_elementary_2012,bauschke_convex_2017} that a maximizing solution exists. Further, using convexity-preserving operations~\cite{boyd_convex_2023}, we can show that all utilized functions $h$, $q$ are proper, closed and convex and the fact that the input constraint set $[0,1]$ is also convex, we know that the optimization problem~\eqref{eqs:optprbustar} is a convex optimization problem.

The fact that this is a convex optimization problem allows us to apply Fermat's rule~\cite{rockafellar_convex_1997}, leading to the equivalent optimality condition for~\eqref{eqs:optprbustar}
\begin{align}
  \begin{aligned}
    &\exists s \in \partial q(u^\ast(\tau)), \; \exists \eta \in \partial \mathcal{I}_{[0,1]}(u^\ast(\tau)) \;:\; \\
    &\qquad\qquad 0 = \mu(\ket{x^\ast(\tau)}, \ket{\lambda^\ast(\tau)})  - \zeta s + \eta \qquad \forall_{\tau \in [0,T]}, \label{eqs:subdiffoptimcond}
  \end{aligned}
\end{align}
where $\mu(\ket{x^\ast(\tau)}, \ket{\lambda^\ast(\tau)}) = \nabla_u h(\ket{x^\ast(\tau)}, \ket{\lambda^\ast(\tau)},  u^\ast(\tau))$ and $\mathcal{I}_{[0,1]}$ denotes the indicator function with respect to $[0,1]$. Further, $\partial f$ denotes the subdifferential of $f$. Note that $\mu$ is independent of $u$ since $h$ is affine in $u$. Concretely, we obtain
\begin{align}
      \mu(\ket{x}, \ket{\lambda})
        &= - \di \bra{\lambda} \underbrace{[B - C]}_F \ket{x} + \di \bra{x} [B - C] \ket{\lambda} \\\nonumber 
        &= - \di \braket{\lambda|F|x} + \di \braket{x|F|\lambda}.
\end{align}

We now determine the optimal input via a case distinction into three cases, corresponding to three values of the subdifferential of $\mathcal{I}_{[0,1]}$:
\begin{enumerate}
  \item The optimal control input is singular, i.e., $u^\ast(\tau)$ lies in the interior of $[0,1]$. Here, the subdifferential is $\partial \mathcal{I}_{[0,1]} = \{0\}$ and~\eqref{eqs:subdiffoptimcond} is equivalent to
  \begin{align}
	  u^\ast(\tau) = {(\partial q)}^{-1}\left( \frac{\mu(\ket{x^\ast(\tau)}, \ket{\lambda^\ast(\tau)})}{\zeta} \right), \label{eqs:analytcustar}
  \end{align}
  assuming the inverse mapping ${(\partial q)}^{-1}$ is unique. 
  Uniqueness holds if $q(u)$ is strictly convex~\cite{rockafellar_convex_1997}, which can be guaranteed if $q(u)=\norm{ H(u) }_F$, i.e., using the Frobenius norm. Studying the uniqueness of ${(\partial q)}^{-1}$ for the spectral norm $q(u)=\norm{H(u)}_2$ is an interesting future research direction. In Section~\ref{sec:simulations}, we show that ${(\partial q)}^{-1}$ is indeed unique when using the spectral norm in a numerical example.
  \item $u^\ast(\tau) = 0$, i.e., it lies on the boundary of $[0,1]$. Here, the subdifferential is $\eta \in \partial \mathcal{I}_{[0,1]} = \{ y \in \R \,|\, y \leq 0 \}$ and~\eqref{eqs:subdiffoptimcond} is equivalent to
  \begin{align}
    \exists_{s \in \partial q(0)} 0 &\geq \eta = \mu(\ket{x^\ast(\tau)}, \ket{\lambda^\ast(\tau)}) - \zeta s \\
     \Leftrightarrow \qquad\qquad \zeta s &\geq \mu(\ket{x^\ast(\tau)}, \ket{\lambda^\ast(\tau)}).
  \end{align}

  The optimal input being zero can be detected via a switching function
  \begin{align}
    \mu(\ket{x^\ast(\tau)}, \ket{\lambda^\ast(\tau)}) - \zeta \max_{s\in\partial q(0)} s \leq 0. \label{eqs:switchingfnc1}
  \end{align}
  \item $u^\ast(\tau) = 1$, i.e., it lies on the boundary of $[0,1]$. Here, the subdifferential is $\eta \in \partial \mathcal{I}_{[0,1]} = \{ y \in \R \,|\, y \geq 0 \}$ and~\eqref{eqs:subdiffoptimcond} is equivalent to
  \begin{align}
      \exists_{s \in \partial q(1)} 0 &\leq \eta = \mu(\ket{x^\ast(\tau)}, \ket{\lambda^\ast(\tau)}) - \zeta s \\
        \Leftrightarrow \qquad\qquad \zeta s &\leq \mu(\ket{x^\ast(\tau)}, \ket{\lambda^\ast(\tau)}).
  \end{align}
  
  The optimal input being one can be detected via a switching function
  \begin{align}
    \mu(\ket{x^\ast(\tau)}, \ket{\lambda^\ast(\tau)}) - \zeta \min_{s\in\partial q(1)} s \geq 0. \label{eqs:switchingfnc2}
  \end{align}
\end{enumerate}

\begin{remark}
    Note here, that all the derivations above are also valid when using any compact, convex input constraint set $\mathcal{U}$ instead of the interval $[0,1]$.
\end{remark}

\paragraph{Singular control input.}\label{lem:singularcontrol}
  Combining the above derivation, the control input $u^\ast(\tau)$ is singular, i.e., it has an annealing structure with smooth variation in $[0,1]$, if the following holds
  \begin{equation}
    \zeta \max_{s \in \partial q(0)} s \leq \mu(\ket{x^\ast(\tau)}, \ket{\lambda^\ast(\tau)}) \leq \zeta \min_{s \in \partial q(1)} s. \label{eqs:switchingfnc}
  \end{equation}

On the other hand, when~\eqref{eqs:switchingfnc} is not fulfilled, the control input is in a bang section, i.e., it is either zero or one.
The case without regularization, i.e., $\zeta=0$, is also included in this result. Setting $\zeta = 0$ yields the singular region $\mu(\ket{x^\ast},\ket{\lambda^\ast}) = 0$, showing that the singular control section in the unregularized case is smaller compared to the robust regularized setup. Also, note that this result is consistent with~\cite{brady_optimal_2021}. 

\begin{remark}
  By leveraging the global phase, it is possible to use an improved robustness measure $\tilde{q}(u) = \min_{\varphi \in \R}  \norm{ H(u) + \varphi \Iden }_2$, see \ref{apx:globalphase} for details. Since $\tilde{q}(u)$ is also convex, the above derivation can be carried out analogously without any modifications.
\end{remark}

In the following, we use the condition~\eqref{eqs:switchingfnc} in order to derive a more insightful, sufficient condition for the absence of a bang-bang section in the optimal control trajectory, i.e., a condition such that $u(\tau)$ is singular for all $\tau \in [0,T]$. To this end, we define the following constants
\begin{align}
    M_{\mathrm{ub}} &= \min_{s \in \partial q(1)} s, \qquad
    M_{\mathrm{lb}} = \max_{s \in \partial q(0)} s.
\end{align}

Further, we have $M_{\mathrm{lb}} \leq M_{\mathrm{ub}}$ since the subdifferential operator of a convex function is a monotone operator \cite{bauschke_convex_2017}. Using the switching function~\eqref{eqs:switchingfnc}, $\norm{\ket{x}}_2 = 1$ and $\norm{\ket{\lambda}}_2 \leq \sigma_\mathrm{max}(C)$ (due to the boundary condition, i.e., $\ket{\lambda^\ast(T)}+C\ket{x^\ast(T)} = 0$), we obtain the following sufficient condition for the absence of a bang-bang section, i.e., a pure annealing solution
\begin{align}
  \zeta M_{\mathrm{lb}} &\leq - 2 \sigma_{\mathrm{max}} (F) \sigma_{\mathrm{max}} (C) \\
  \textrm{and} \qquad \zeta M_{\mathrm{ub}} &\geq + 2 \sigma_{\mathrm{max}} (F) \sigma_{\mathrm{max}} (C) \label{eqs:absbangbangcond},
\end{align}
where $\sigma_\mathrm{max}(\cdot)$ denotes the maximum singular value.

To summarize, we have shown that, if equation~\eqref{eqs:absbangbangcond} holds, the entire input trajectory is singular.  We can summarize equation~\eqref{eqs:absbangbangcond} as the condition
\begin{align}
  \zeta \geq 2 \frac{\sigma_{\mathrm{max}} (F) \sigma_{\mathrm{max}}(C)}{\min \{ | M_{\mathrm{lb}} |, | M_{\mathrm{ub}} | \}}
\end{align}
assuming that $M_{\mathrm{lb}}$ is negative and $M_{\mathrm{ub}}$ is positive. If $M_{\mathrm{ub}}$ and $M_{\mathrm{lb}}$ do share the same sign, there is always a bang section but not necessarily a bang-bang section, i.e., the optimal solution may be zero or one at the start or at the end, but not necessarily at both.

\begin{remark}
    If $M_{\mathrm{ub}}$ and $M_{\mathrm{lb}}$ share the same sign, this intuitively means the global minimum of $q$ lies outside of the range $[0,1]$.
\end{remark}

In summary, our theoretical results imply that the optimal solution of the robust optimal annealing problem is substantially different from ideal quantum (i.e., not robust) annealing, leading to a larger singular region. This indicates that continuously varying inputs (as in QA) admit superior robustness against coherent control errors in comparison to pure bang-bang solutions (as in QAOA). 

Intuitively, increasing the singular region leads to a smaller norm in the overall applied Hamiltonian. Since coherent control errors are multiplicative, this means that the errors impact the trajectory less, hence increasing the robustness against this type of error signal.

\section{Numerical simulations}\label{sec:simulations}

\begin{figure}
  \centering
  \begin{subfigure}[b]{0.49\textwidth}\centering
    \includegraphics[]{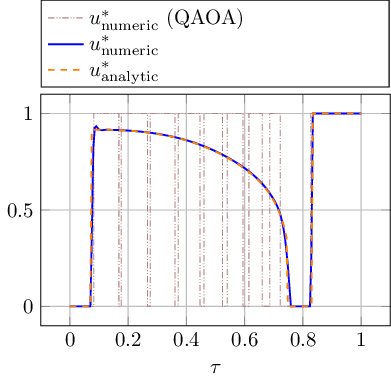}
    \caption{input $u^\ast$}\label{fig:exmp_x8_normnone_u}
  \end{subfigure}
  \hfill
  \begin{subfigure}[b]{0.49\textwidth}\centering
    \includegraphics[]{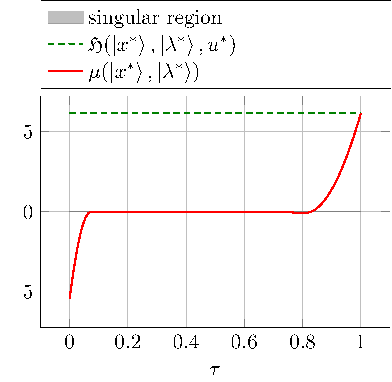}
    \caption{switching function}\label{fig:exmp_x8_normnone_H}
  \end{subfigure}
  \caption[Example: 8-qubit; $\zeta = 0$; nominal case]{Example: 8-qubit; $\zeta = 0$; nominal case, i.e., classical QA as in Section~\ref{prb:optimalcontrol}. This Figure shows the optimal annealing protocol and the conditions for a singular control section. In addition, the corresponding control Hamiltonian $\chamil$ is plotted.}\label{fig:exmp_x8_normnone}
\end{figure}

\begin{figure}
  \centering
  \begin{subfigure}[h]{0.49\textwidth}\centering
    \includegraphics[]{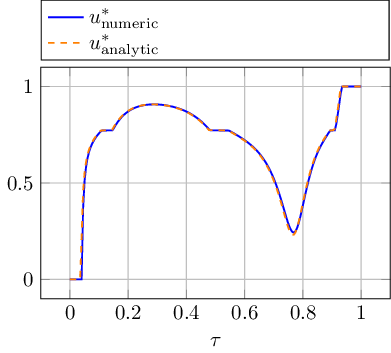}
    \caption{input $u^\ast$}\label{fig:exmp_x8_norm2_zeta02_u}
  \end{subfigure}
  \hfill
  \begin{subfigure}[h]{0.49\textwidth}\centering
    \includegraphics[]{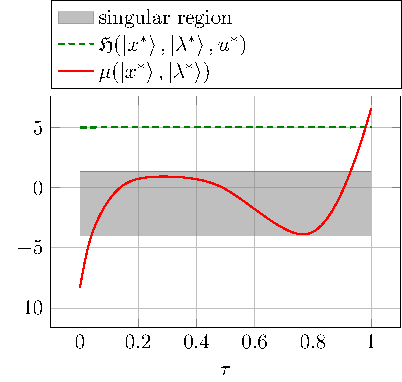}
    \caption{switching function}\label{fig:exmp_x8_norm2_zeta02_H}
  \end{subfigure}
  \caption[Example: 8-qubit; $\zeta = 0.1$; spectral norm]{Example: 8-qubit; $\zeta = 0.1$; spectral norm regularization, i.e., $q(u) = \norm{H(u)}_2$. This Figure shows the robust optimal annealing protocol and the conditions for a singular control section. In addition, the corresponding control Hamiltonian $\chamil$ is plotted.}\label{fig:exmp_x8_norm2_zeta02}
\end{figure}

\begin{figure}
  \centering
  \begin{subfigure}[b]{0.49\textwidth}\centering
    \includegraphics[]{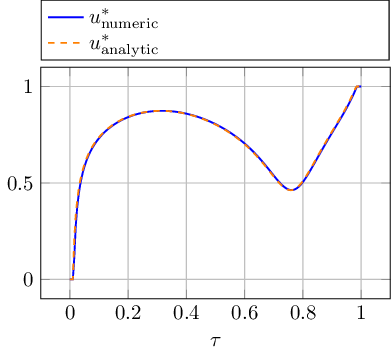}
    \caption{input $u^\ast$}\label{fig:x8params_normF_zeta01_u}
  \end{subfigure}
  \hfill
  \begin{subfigure}[b]{0.49\textwidth}\centering
    \includegraphics[]{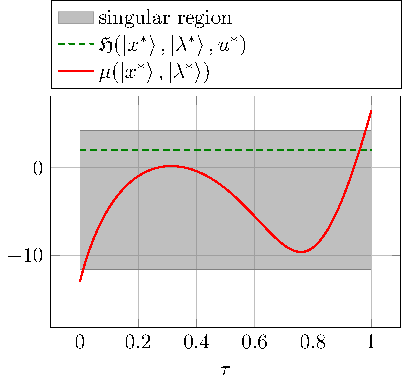}
    \caption{switching function}\label{fig:x8params_normF_zeta01_H}
  \end{subfigure}
  \caption[Example: 8-qubit; $\zeta = 0.1$; Frobenius norm]{Example: 8-qubit; $\zeta = 0.1$; Frobenius norm as regularization, i.e., $q(u) = \norm{H(u)}_F$. This Figure shows the robust optimal annealing protocol and the conditions for a singular control section. In addition, the corresponding control Hamiltonian $\chamil$ is plotted.}\label{fig:x8params_normF_zeta01}
\end{figure}

Now, we turn our attention to numerical simulation in order to shed some light on the theoretical results. We solve the robust optimal QA problem in Section~\ref{prb:optimalcontrol} numerically using MATLAB~\cite{MATLAB}. The simulation results for a randomly generated Ising model are shown in Figures~\ref{fig:exmp_x8_normnone},~\ref{fig:exmp_x8_norm2_zeta02} and~\ref{fig:x8params_normF_zeta01}, i.e., for a random choice of the matrix $J$ in equation~\eqref{eqs:isingmodel}. We discretize the time span $[0,T]$ and use gradient descent in combination with analytically computed gradients to optimize for the best QA protocols. 
In general, it is not guaranteed that the numerically found solution is globally optimal. Thus, to reduce the risk of ending up in a local minimum, we reoptimized the QA protocol using different initial conditions and parameters in the gradient decent algorithm, consistently leading to the same optimal protocol. In the case of QAOA, we use the time intervals as optimization variables. The source code can be found online\footnote{
\url{https://github.com/eragon10/njp_2024_roqa}
}.

Figure~\ref{fig:exmp_x8_normnone_u} shows the nominal optimal input protocol, i.e., without regularization $\zeta = 0$, and the optimal input protocol found using the QAOA approach without regularization~\cite{brady_optimal_2021}. In Figure~\ref{fig:exmp_x8_normnone_H}, we see the corresponding switching function~\eqref{eqs:switchingfnc}. As expected based on the maximum principle, compare~\eqref{eqs:switchingfnc} and~\cite{brady_optimal_2021},  $\mu(\ket{x^\ast}, \ket{\lambda^\ast})$ is zero in the interval of the singular control section. Further, it starts and ends with a bang as expected. 
Moreover, the control Hamiltonian $\mathfrak{H}$ must be constant over time, which is another necessary optimality condition in Pontryagin's maximum principle. We use the QAOA and nominal protocols as reference protocols to compare them with the robustness-enhanced versions.

Next, Figure~\ref{fig:exmp_x8_norm2_zeta02_u} shows the robust optimal annealing protocol which is computed by solving the robust quantum optimal control problem in Section~\ref{sec:robust_annealing} with regularization parameter $\zeta = 0.2$. The singular region is an interval in the robust case, compare Section~\ref{lem:singularcontrol}. Hence, we obtain a larger singular control section with smoothly varying input compared to the nominal case. As long as $\mu(\ket{x^\ast},\ket{\lambda^\ast})$ lies inside the singular region, the optimal input $u^{\ast}$ is singular. However, the optimal QA protocol still starts and ends with a bang. The kinks of the optimal input $u^{\ast}$ can be explained by the non-smoothness of the spectral norm.

The optimal input $u^\ast_\mathrm{analytic}$, computed via~\eqref{eqs:analytcustar} using the numerically found optimal solution, i.e., state $\ket{x^\ast}$ and co-state $\ket{\lambda^\ast}$, coincides with the numerically found solution and one can check that $q(u)$ is strictly convex in the interval $[0,1]$ for this numerical example. Hence, the inverse mapping ${(\partial q)}^{-1}(\cdot)$ is unique.

Recall that our results apply analogously to using different matrix norms as regularization. We now consider the Frobenius norm, which is differentiable and, therefore, yields smoother results. Figure~\ref{fig:x8params_normF_zeta01_u} confirms this. There, we see the optimal protocols corresponding to a Frobenius norm regularization $q(u) = \norm{H(u)}_F$ with parameter $\zeta = 0.1$. 

In closing, we compare the robustness, i.e., worst-case fidelity, between the different optimal annealing protocols over increasing noise levels of $\hat{\epsilon}$. We use a set of 20 randomly generated error signals, i.e., discretize the error signal in 20 sections and draw the amplitude of each discretization step uniformly at random from a uniform distribution, for our analysis. Then, we scaled the error signals according to $\hat{\epsilon}$. For each value of $\hat{\epsilon}$, we simulated each of the four approaches and plotted the worst-case fidelity over all noise signals.

\begin{figure}[!ht]
  \begin{center}
    \includegraphics[]{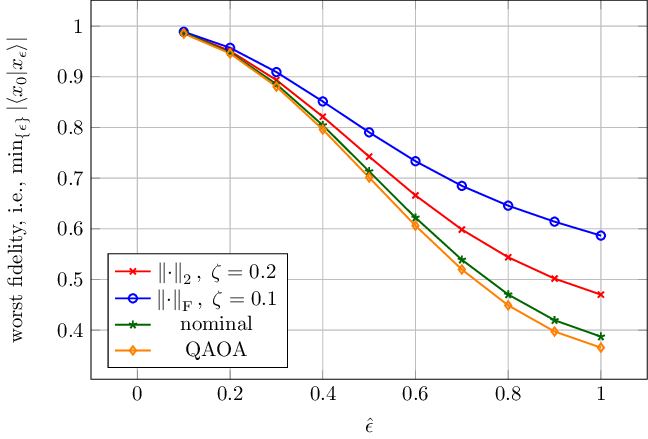}
  \end{center}
  \caption[Example: 8-qubit; worst fidelity]{Example: 8-qubit. Worst fidelity the optimal annealing protocols over $\hat{\epsilon}$ corresponding to the four approach nominal, QAOA, spectral norm, Frobenius norm. The worst fidelity describes the worst fidelity in the set of generated error signals scaled according to $\hat{\epsilon}$.}\label{fig:rob_x8_fidelity}
\end{figure}

Figure~\ref{fig:rob_x8_fidelity} shows that, for each of the approaches, the fidelity decreases with increasing $\hat{\epsilon}$. Moreover, the two robustness-enhanced protocols, i.e., robust optimal annealing with regularization based on the spectral norm and Frobenius norm, yield a higher fidelity and deteriorate more slowly than the nominal and QAOA protocols.

\begin{figure}[!ht]
  \begin{center}
    \includegraphics[]{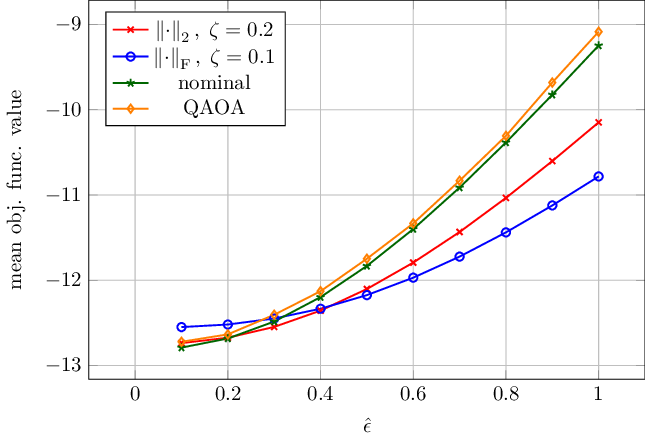}
  \end{center}
  \caption[Example: 8-qubit; Mean objective function value.]{Example: 8-qubit. Mean objective function value, i.e., equation~\eqref{eqs:objectivefunction}. We are using the set of generated error signals to compute a mean value of $\braket{x_\epsilon(T)|C|x_\epsilon(T)}$.}\label{fig:rob_x8_objvalue}
\end{figure}

Finally, we compare the cost values obtained using the different strategies in Figure~\ref{fig:rob_x8_objvalue}. Due to the additional regularization, the optimal value of the robust approaches is larger for small noise levels. However, the objective function value increases more slowly with increasing $\hat{\epsilon}$ in the robustness-enhanced versions than in the standard QA and QAOA setups, indicating superior robustness due to the regularization.

\begin{figure}[!ht]
  \centering
  \begin{subfigure}[h]{0.49\textwidth}\centering
    \includegraphics[]{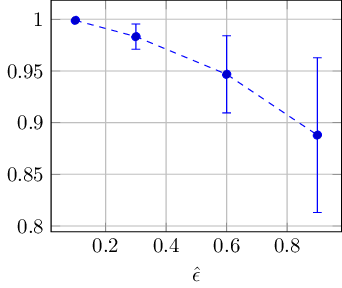}
	\caption{worst-fidelity ration between robust and nominal QA protocols}\label{fig:rob_x6_rndising_fidelity}
  \end{subfigure}
  \hfill
  \begin{subfigure}[h]{0.49\textwidth}\centering
    \includegraphics[]{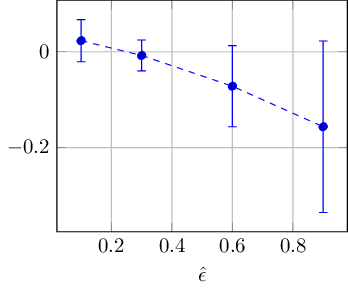}
	\caption{relative cost difference between robust and nominal QA protocols}\label{fig:rob_x6_rndising_objfncval}
  \end{subfigure}
  \caption[Example: 6-qubit; Randomly generated Ising Models.]{Example: 6-qubit; 250 randomly generated Ising models. Figure~\ref{fig:rob_x6_rndising_fidelity} shows the mean value and standard deviation of the ratio between the robust and nominal fidelity~\eqref{eqs:ratio_fidelity} for the 250 randomly generated Ising models and for different noise levels $\hat{\epsilon}$. Similarly, Figure~\ref{fig:rob_x6_rndising_objfncval} displays the relative cost difference~\eqref{eqs:ratio_objfncval}.}\label{fig:rob_x6_rndising}
\end{figure}

In addition, we demonstrate the broad applicability of our theoretical results by considering $250$ different randomly generated Ising models with $6$ qubits. Figure~\ref{fig:rob_x6_rndising} shows that we obtain similar results as for the single Ising model above.
For each randomly generated Ising model, we numerically computed the corresponding optimal nominal and robust QA protocols. 

Figure~\ref{fig:rob_x6_rndising} compares the worst-case fidelity for robust quantum annealing and for nominal quantum annealing, denoted by $\mathsf{f}_\mathrm{rob}$ and $\mathsf{f}_\mathrm{nom}$, via their ratio 
\begin{equation}
	\frac{\mathsf{f}_\mathrm{rob}}{\mathsf{f}_\mathrm{nom}} \label{eqs:ratio_fidelity}
\end{equation}
depending on the size of the error $\hat{\epsilon}$. Further, the figure compares the averaged robust cost $J_\mathrm{rob}$ and the nominal cost $J_\mathrm{nom}$ via the relative difference
\begin{equation}
	\frac{ J_\mathrm{rob} - J_\mathrm{nom} }{ J_\mathrm{nom}(\hat{\epsilon} = 0.1) } \label{eqs:ratio_objfncval}
\end{equation}
Both the fidelity ratio and the cost difference are computed for all 250 Ising models, and the figure displays the resulting average value as well as the confidence interval based on the standard deviation.

Note that the fidelity ratio in Figure~\ref{fig:rob_x6_rndising_fidelity} is consistently below $1$, confirming the improved robustness of robust quantum annealing. Similarly, for Figure~\ref{fig:rob_x6_rndising_objfncval}, values below $0$ indicate that the robust QA protocol yields a better objective function value. As expected, for small error signals, the nominal QA protocol performs better, but with increasing error, the robust QA protocol leads to superior performance.

To summarize, in the considered example, the proposed robust optimal QA protocols provide substantial cost improvements over classical QA in the presence of coherent control errors. Further, as predicted by the theoretical results in Section~\ref{sec:robust_optimal_annealing}, the robust protocols indeed admit a larger singular region in comparison to the nominal quantum optimal control problem without regularization.

\section{Conclusion}\label{sec:conclusion}
In this paper, we studied the robustness of optimal annealing protocols against coherent control errors.
We showed that the robustness is quantified by the norm of the Hamiltonian, indicating that solutions with smaller norms are more robust.
Motivated by this result, we proposed a robust QA approach which minimizes the sum of two terms: the cost function and a regularization term depending on the norm of the Hamiltonian.
We studied the optimal solution of the corresponding optimal control problem, showing that the singular region is substantially enlarged.
This suggests that singular control inputs admit improved robustness over bang-bang solutions, indicating that QA solutions are more robust than QAOA.
Our theoretical findings are confirmed in simulation.

Our results open the door for designing and studying QA protocols and variational quantum algorithms~\cite{cerezo2021variational} which not only produce good results in an ideal, noise-free setup but also work reliably in the presence of error occurring on real-world quantum devices.
In particular, it will be interesting to extend our findings to different quantum optimization techniques as well as different error models, showing for a variety of setups that considering robustness during the optimization may lead to more reliable results.
To this end, different classes of coherent errors such as coherent phase errors~\cite{bravyi2018correcting}, which are described by additional Pauli $Z$ rotations with unknown angle, would be a natural starting point.
  Moreover, the effect of decoherence is of particular importance on current quantum devices.
  The robustness against decoherence may be a conflicting objective to the robustness against coherent control errors pursued in the present paper:
  The presented robust optimal quantum annealing protocols result from a regularized cost which, roughly speaking, reduces the strength of the Hamiltonian.
  This may lead to a possible increase of the transition time to the ground state, hence increasing the damage of decoherence. 
  Studying this conflict in more detail and designing protocols which are robust against both types of errors, e.g., via optimal control with free end time, is an interesting future research direction.

\section*{Acknowledgement}
This work was funded by Deutsche Forschungsgemeinschaft (DFG, German Research Foundation) under Germany’s Excellence Strategy - EXC 2075 - 390740016. We acknowledge the support by the Stuttgart Center for Simulation Science (SimTech).

\printbibliography[heading=bibintoc]

\appendix

\section{Technical assumptions}\label{apx:assumptions}

If the right-hand side of a differential equation is globally Lipschitz continuous a unique solution of this differential equation exists for any initial condition~\cite{khalil_nonlinear_2002}. To ensure this, let the following assumptions be satisfied.
\begin{enumerate}[itemindent=2em,leftmargin=2em,start=1,label={(A\arabic*):}]
  \item The Hamiltonian $H(u(\tau))$ is bounded for all times, i.e.,\label{asm:inner_rob_1}
  \begin{align}
    \exists M \in \R_{+} \,:\, \forall_{\tau \in [0,T]} \; \norm{ H(u(\tau)) }_2 \leq M.
  \end{align}
  This is satisfied if $\forall_{\tau \in [0,T]} : u(\tau) \in [0,1]$ and 
  \begin{equation}
    \exists M \in \R_+ \,:\, \forall_{u \in [0,1]} \; \norm{H(u)}_2 \leq M.
  \end{equation}
  \item The Hamiltonian $H(u(\tau))$ has at most a countably infinite number of discontinuities.
  \item The error signal $\epsilon(\tau)$ is bounded for all times, i.e.,\label{asm:inner_rob_3}
  \begin{align}
    \exists B \in \R_+ \,:\, \forall_{\tau \in [0,T]}\; \abs{ \epsilon(\tau) } \leq B.
  \end{align}
  \item The error signal $\epsilon(\tau)$ has at most a countably infinite number of discontinuities.\label{asm:inner_rob_4}
\end{enumerate}

\section{Derivation of the robustness bound~\eqref{eqs:lipboundfidelity}}\label{apx:thmrobustbound}

The proof follows the lines of~\cite{berberich_robustness_2023} and is included in the following for completeness. Let assumptions of~\ref{apx:assumptions} hold and observe that
\begin{equation}
	\begin{aligned}
	\norm{\ket{x_\epsilon(T)} - \ket{x_0(T)}}^2 &= \braket{x_\epsilon(T) | x_\epsilon(T)} - \braket{x_0(T) | x_\epsilon(T)} \\
    & \hspace{1.4cm} - \braket{x_\epsilon(T) | x_0(T)} + \braket{x_0(T) | x_0(T)} \\
	&= 2 - 2 \Re{\braket{x_\epsilon(T)|x_0(T)}} \\
	&\geq 2 - 2 \abs{\braket{x_\epsilon(T) | x_0(T)}}.
	\end{aligned}
\end{equation}
Hence,
\begin{equation}
  2 - 2 \abs{\braket{x_\epsilon(T) | x_0(T)}} \leq L^2 \hat{\epsilon}^2.
\end{equation}

Note here that $\sup_{\tau \in [0,T]} \abs{\epsilon'(\tau)-\epsilon(\tau)}$ is finite by assumption. Hence, if we can compute an upper bound of the Lipschitz bound $L$, we can directly conclude that the fidelity is bounded as
\begin{equation}
  \abs{\braket{x_\epsilon(T)|x_0(T)}} \geq 1 - \frac{L^2 \hat{\epsilon}^2}{2}.
\end{equation}

\section{Proof that~\eqref{eqs:lipschitzupperbound} is a Lipschitz bound}\label{apx:thmlipschitzbound}

In the following, we view the Hamiltonian $H(u)$ as a function of $\tau$, i.e., $H(u(\tau)) = H(\tau)$. Hence, we prove that for any initial condition $\ket{x^0}$ and any QA protocol $u(\tau) : [0;T] \to [0,1]$ the following holds
\begin{align}
   \norm{\ket{x_{\epsilon'}(T)} - \ket{x_{\epsilon}(T)}}_2 \leq \underbrace{ \int_0^T \norm{H(\tau)}_2 \,\dd \tau }_{L}  \norm{\epsilon' - \epsilon}_{\supnorm} \qquad \forall_{T \geq 0}. \label{eqs:apxlipschitzbound}
\end{align}

By assumption, the system~\eqref{eqs:coherentcontrolerr} possesses a unique solution. We describe the solution of~\eqref{eqs:coherentcontrolerr} using a state transition matrix $\Phi_\epsilon(t_1, t_0)$, i.e.,
\begin{align}
  \ket{x_\epsilon(t_1)} = \Phi_\epsilon(t_1, t_0) \ket{x_\epsilon(t_0)}.
\end{align}

The state transition matrix possesses the following properties~\cite{baake_peano-baker_2011}
\begin{enumerate}
  \item $\frac{\dd}{\dd \tau} \Phi_\epsilon(\tau,\cdot) = A_\epsilon(\tau) \Phi_\epsilon(\tau, \cdot)$
  \item $\Phi_\epsilon(t_1, t_0) = \Iden + \int_{t_0}^{t_1} A_\epsilon(\tau) \Phi_\epsilon(\tau, t_0) \; \dd \tau$
\end{enumerate}
where $A_\epsilon(\tau)= - \di (1+\epsilon(\tau)) H(\tau)$.  Using our findings, we can reformulate the Lipschitz bound condition, i.e., equation~\eqref{eqs:apxlipschitzbound}, with respect to $\ket{x_\epsilon}$ into one with respect to $\Phi_\epsilon$, i.e.,
\begin{align}
    \norm{ \ket{x_{\epsilon'}(T)} - \ket{x_\epsilon(T)} }_2 &= \norm{ \Phi_{\epsilon'}(T,0) \ket{x^0} - \Phi_\epsilon(T,0) \ket{x^0} }_2 \\
    &= \norm{ \left( \Phi_{\epsilon'}(T,0) - \Phi_\epsilon(T,0) \right) \ket{x^0} }_2 \\
    &\leq \norm{ \Phi_{\epsilon'}(T,0) - \Phi_\epsilon(T,0) }_2
\end{align}
using the fact that $\norm{\ket{x^0}}_2 = 1$. Hence, it suffices to find a Lipschitz bound for the state transition matrix.

Assume the state transition matrix $\Phi_\epsilon$ is Gateaux differentiable with respect to $\epsilon$. Then, applying the generalized mean value theorem~\cite{pathak_introduction_2018} yields
\begin{align}
  \norm{ \Phi_{\epsilon'} (T, 0) - \Phi_{\epsilon} (T, 0) }_2 \leq \sup_{\beta \in (0, 1)} \norm{ \gateaux_{\epsilon' - \epsilon} \Phi_{\epsilon + \beta (\epsilon' - \epsilon)} (T, 0) }_2, \label{eqs:apxmeanvaluethm}
\end{align}
where $\gateaux_h F(x) = \lim_{\beta \to 0} \frac{ F(x+ \beta h) - F(x)}{\beta}$ is the Gateaux derivative~\cite{pathak_introduction_2018}.

Using the properties of the state transition matrix, we obtain
\begin{align}
       \gateaux_{\delta \epsilon} \Phi_{\epsilon} (t, 0)
          &= \lim_{\beta \to 0} \frac{\Phi_{\epsilon + \beta \delta  \epsilon} (t, 0) - \Phi_{\epsilon} (t, 0)}{\beta} \\
          &= \lim_{\beta \to 0} \frac{\Iden - \di \int_0^t (1 + \epsilon(\tau) + \beta \delta \epsilon(\tau)) H(\tau) \Phi_{\epsilon + \beta \delta  \epsilon} (\tau, 0) \;\dd \tau }{\beta} \\
          & \qquad\qquad + \frac{- \Iden + \di \int_0^t (1 + \epsilon(\tau)) H(\tau) \Phi_{\epsilon} (\tau, 0) \;\dd  \tau
          }{\beta}  \\
          &= \lim_{\beta \to 0} \frac{-\di \int_0^t (1 + \epsilon(\tau)) H(\tau) \left[\Phi_{\epsilon + \beta \delta \epsilon} (\tau, 0) - \Phi_{\epsilon} (\tau, 0)\right] \;\dd \tau }{\beta} \\
          & \qquad\qquad + \frac{- \di \beta \int_0^t \delta  \epsilon(\tau) H(\tau) \Phi_{\epsilon + \beta \delta \epsilon} (\tau, 0) \;\dd \tau 
          }{\beta} \\
          &= - \di \int_0^t (1 + \epsilon(\tau)) H(\tau) \underbrace{\gateaux_{\delta \epsilon} \Phi_{\epsilon} (\tau,0)}_{q (\tau)} \;\dd \tau \\ &\hspace{3cm} - \di \int_0^t \delta \epsilon(\tau) H(\tau)  \Phi_{\epsilon} (\tau, 0) \;\dd \tau.
\end{align}

For $q(t) =  \gateaux_{\delta \epsilon} \Phi_{\epsilon} (t, 0)$, this implies
\begin{align}
  q (t) &= - \di \int_0^t (1 + \epsilon(\tau)) H (\tau) q (\tau) \;\dd \tau \\ &\hspace{3cm} - \di \int_0^t \delta \epsilon (\tau) H(\tau) \Phi_{\epsilon} (\tau, 0) \;\dd \tau. \label{eqs:apxsensitivityode}
\end{align}

We can now differentiate~\eqref{eqs:apxsensitivityode} to get a differential equation in terms of $q(\tau)$
\begin{align}
  \dot{q}(t) &=  \underbrace{- \di (1 + \epsilon (t)) H (t) }_{A (t)} q(t) \underbrace{- \di \delta \epsilon (t) H (t) \Phi_{\epsilon} (t, 0)}_{b (t)} \\
  \dot{q} (t) &= A (t) q (t) + b (t)
  \label{eqs:errboundodegder}
\end{align}
with initial condition $q_0 = 0 \in \C^{d \times d}$. Due to the assumptions in~\ref{apx:assumptions}, this differential equation also satisfies the condition of global existence and uniqueness. Considering only the homogeneous part $\dot{q}(t)=A(t)q(t)$, note that the state transition matrix of~\eqref{eqs:errboundodegder} and~\eqref{eqs:coherentcontrolerr} coincide. Hence, the solution is given by~\cite{baake_peano-baker_2011}
\begin{align}
    \gateaux_{\delta \epsilon} \Phi_{\epsilon} (T, 0) = q (T) &= \Phi_{\epsilon} (T, 0) q_0 + \int_0^T \Phi_{\epsilon} (T, \tau) b (\tau) \;\dd \tau \\
    &= - \di \int_0^T \delta \epsilon (\tau) \Phi_{\epsilon} (T,\tau) H (\tau) \Phi_{\epsilon} (\tau, 0) \;\dd \tau. \label{eqs:apxsensitivitysol}
\end{align}

Recall that we assume Gateaux differentiability. This is indeed the case since the existence and uniqueness of the solution of~\eqref{eqs:errboundodegder} is guaranteed~\cite{khalil_nonlinear_2002}.

Now, inserting~\eqref{eqs:apxsensitivitysol} into~\eqref{eqs:apxmeanvaluethm}, yields
\begin{align}
    &\norm{ \Phi_{\epsilon'} (T, 0) - \Phi_{\epsilon} (T, 0) }_2 \\
      \leq& \sup_{\beta \in (0, 1)} \norm{ \gateaux_{\epsilon' - \epsilon} \Phi_{\epsilon + \beta (\epsilon' - \epsilon)} (T, 0) }_2 \\
      =& \sup_{\beta \in (0, 1)} \norm{ - \di \int_0^T (\epsilon'(\tau) - \epsilon (\tau)) \Phi_{\epsilon + \beta (\epsilon' -  \epsilon)} (T, \tau) \right. \\ &\hspace{4cm} \left. H (\tau) \Phi_{\epsilon + \beta (\epsilon' -\epsilon)} (\tau, 0) \;\dd \tau \vphantom{\int_0^T} }_2 \\
      \leq& \sup_{\beta \in (0, 1)} \int_0^T \abs{ \epsilon'(\tau) - \epsilon(\tau) } \norm{ \Phi_{\epsilon + \beta (\epsilon' - \epsilon)} (T, \tau) }_2 \\ &\hspace{4cm} \norm{ H (\tau) }_2  \norm{ \Phi_{\epsilon + \beta (\epsilon' - \epsilon)} (\tau, 0) }_2 \;\dd \tau \\
      \leq&  \left(\sup_{\tau \in [0, T]} \abs{ \epsilon'(\tau) - \epsilon(\tau) } \right)  \int_0^T \norm{ H(\tau) }_2 \;\dd \tau \\
      =& \hat{\epsilon}  \int_0^T \norm{ H(\tau) }_2 \;\dd \tau.
\end{align}

In addition, we are using the fact that the state transition matrix is unitary.

\section{Extension: Global phase}\label{apx:globalphase}

For our analysis, the robustness bound~\eqref{eqs:apxlipschitzbound} can be improved by leveraging the fact that a global phase $y = \de^{- \di \beta} x$ does not appear when observing quantum states, e.g., 
\begin{align}
  \braket{y | C | y} &= \braket{\de^{ -\di \beta} x | C | \de^{- \di \beta} x} = \braket{x | C | x}.
\end{align}

The same holds for the fidelity of a quantum state, i.e., $\abs{\braket{\cdot|y}} = \abs{\braket{\cdot|x}}$. More precisely, the family of Hamiltonians
\begin{align}
  H_{\varphi}(\tau) = H(\tau) + \varphi(\tau) \Iden
\end{align}
all generate the same unitary state transition matrix modulo a global phase. We claim that
\begin{align}
  \de^{- \di \int_0^t \varphi (\tau) \;\dd \tau} \Phi_0 (t, 0) = \Phi_{\varphi} (t, 0)
\end{align}
where $\Phi_{\varphi} (t, \cdot)$ denotes the state transition matrix of the linear time-varying differential equation $\ket{\dot{x}} = - \di (H(\tau) + \varphi(\tau) \Iden) \ket{x}$. That means that adding the term $\varphi(\tau) \Iden$ to the system Hamiltonian causes a global phase on the transition matrix $\Phi(t,0)$. Hence, a possibly improved error bound is
\begin{align}
  \norm{\ket{x_{\epsilon'}(T)} - \ket{x_{\epsilon}(T)}}_2 \leq  \int_0^T \norm{H(\tau) + \varphi(\tau) \Iden} \;\dd \tau \norm{\epsilon' - \epsilon}_{\supnorm}.
\end{align}

To prove that, recall that $\de^{- \di 0} \Phi_0 (0, 0) = \Phi_{\varphi}(0, 0) = \Iden$. Since
\begin{align}
    \frac{\dd}{\dd t} &\underbracket{\left[ \de^{- \di \int_0^t \varphi
    (\tau) \;\dd \tau} \Phi_0 (t, 0) - \Phi_{\varphi} (t, 0) \right]}_{\delta} \\
      &= \de^{- \di \int_0^t \varphi (\tau) \;\dd \tau} \left[- \di \varphi \Phi_0 (t, 0) - \di H \Phi_0 (t, 0) \right] + \di (H + \varphi \Iden) \Phi_{\varphi} (t, 0) \\
      &= - \di (H + \varphi \Iden) \de^{- \di \int_0^t \varphi (\tau) \;\dd \tau} \Phi_0 (0, t) + \di (H + \varphi \Iden) \Phi_{\varphi} (t, 0) \\
      &= - \di (H + \varphi \Iden)  \underbracket{\left[ \de^{-\di \int_0^t \varphi (\tau) \;\dd \tau} \Phi_0 (t, 0) - \Phi_{\varphi} (t, 0) \right]}_{\delta},
\end{align}
we have
\begin{align}
  \dot{\delta} = - \di (H + \varphi \Iden) \delta, \quad\quad \delta (0) = 0.
\end{align}
  
Hence, we can conclude that $\delta = 0$ for all times. This conclusion completes our proof that $\de^{-\di \int_0^t \varphi (\tau) \;\dd \tau} \Phi_0 (t, 0) = \Phi_{\varphi} (t,0)$.

This fact can be used to improve the bound on coherent control errors. As $\varphi$ does not change the final objective function value nor the final fidelity, we can drop the identity matrix during the time evolution. Since $\varphi$ is arbitrary, we can minimize over $\varphi(\cdot)$. To sum this up, the improved robustness measure reads as
\begin{equation}
  \tilde{q}(u) = \min_{\varphi \in \R} \norm{H(\tau) + \varphi (\tau)\Iden}_2.
\end{equation}

\end{document}